\pdfoutput=1
\documentclass[12pt,a4paper]{article}
\usepackage{ifthen}
\newboolean{pdflatex}
\setboolean{pdflatex}{true}
\newboolean{articletitles}
\setboolean{articletitles}{true}
\newboolean{uprightparticles}
\setboolean{uprightparticles}{false}
\newboolean{inbibliography}
\setboolean{inbibliography}{false}
\usepackage{microtype}
\usepackage{lineno}  % for line numbering during review
\usepackage{xspace} % To avoid problems with missing or double spaces after
\usepackage{caption} %these three command get the figure and table captions automatically small

\usepackage{graphicx}  % to include figures (can also use other packages)
\usepackage{color}
\usepackage{colortbl}
\graphicspath{{./figs/}} % Make Latex search fig subdir for figures

\usepackage{amsmath} % Adds a large collection of math symbols
\usepackage{amssymb}
\usepackage{amsfonts}
\usepackage{upgreek} % Adds in support for greek letters in roman typeset

\newcommand*\patchAmsMathEnvironmentForLineno[1]{%
\expandafter\let\csname old#1\expandafter\endcsname\csname #1\endcsname
\expandafter\let\csname oldend#1\expandafter\endcsname\csname
end#1\endcsname
 \renewenvironment{#1}%
   {\linenomath\csname old#1\endcsname}%
   {\csname oldend#1\endcsname\endlinenomath}%
}
\newcommand*\patchBothAmsMathEnvironmentsForLineno[1]{%
  \patchAmsMathEnvironmentForLineno{#1}%
  \patchAmsMathEnvironmentForLineno{#1*}%
}
\AtBeginDocument{%
\patchBothAmsMathEnvironmentsForLineno{equation}%
\patchBothAmsMathEnvironmentsForLineno{align}%
\patchBothAmsMathEnvironmentsForLineno{flalign}%
\patchBothAmsMathEnvironmentsForLineno{alignat}%
\patchBothAmsMathEnvironmentsForLineno{gather}%
\patchBothAmsMathEnvironmentsForLineno{multline}%
\patchBothAmsMathEnvironmentsForLineno{eqnarray}%
}

\usepackage{hyperref}    % Hyperlinks in references
\usepackage[all]{hypcap} % Internal hyperlinks to floats.

\def\lhcb {\mbox{LHCb}\xspace}

\def\MagUp {\mbox{\em Mag\kern -0.05em Up}\xspace}

\ifthenelse{\boolean{uprightparticles}}%
{
 
 \def\Pgamma      {\ensuremath{\upgamma}\xspace}

 \def\Ppi         {\ensuremath{\uppi}\xspace}                 
                  
 \def\Prho        {\ensuremath{\uprho}\xspace}

 \def\PDelta      {\ensuremath{\Delta}\xspace}                 
 \def\PXi      {\ensuremath{\Xi}\xspace}                 
 \def\PLambda      {\ensuremath{\Lambda}\xspace}                 
 \def\PSigma      {\ensuremath{\Sigma}\xspace}                 
 \def\POmega      {\ensuremath{\Omega}\xspace}                 
 \def\PUpsilon      {\ensuremath{\Upsilon}\xspace}                 
 
 \def\PB      {\ensuremath{\mathrm{B}}\xspace}                 
                  
 \def\PD      {\ensuremath{\mathrm{D}}\xspace}

 \def\PK      {\ensuremath{\mathrm{K}}\xspace}

 \def\PW      {\ensuremath{\mathrm{W}}\xspace}

 \def\Pb      {\ensuremath{\mathrm{b}}\xspace}                 
 \def\Pc      {\ensuremath{\mathrm{c}}\xspace}

 \def\Pi      {\ensuremath{\mathrm{i}}\xspace}

 \def\Pp      {\ensuremath{\mathrm{p}}\xspace}

 \def\Ps      {\ensuremath{\mathrm{s}}\xspace}

}
{
 
 \def\Pgamma      {\ensuremath{\gamma}\xspace}

 \def\Ppi         {\ensuremath{\pi}\xspace}                 
                  
 \def\Prho        {\ensuremath{\rho}\xspace}

 \mathchardef\PDelta="7101
 \mathchardef\PXi="7104
 \mathchardef\PLambda="7103
 \mathchardef\PSigma="7106
 \mathchardef\POmega="710A
 \mathchardef\PUpsilon="7107
                  
 \def\PB      {\ensuremath{B}\xspace}                 
                  
 \def\PD      {\ensuremath{D}\xspace}

 \def\PK      {\ensuremath{K}\xspace}

 \def\PW      {\ensuremath{W}\xspace}

 \def\Pb      {\ensuremath{b}\xspace}                 
 \def\Pc      {\ensuremath{c}\xspace}

 \def\Pi      {\ensuremath{i}\xspace}

 \def\Pp      {\ensuremath{p}\xspace}

 \def\Ps      {\ensuremath{s}\xspace}

}

\makeatletter
\ifcase \@ptsize \relax% 10pt
  \newcommand{\miniscule}{\@setfontsize\miniscule{4}{5}}% \tiny: 5/6
\or% 11pt
  \newcommand{\miniscule}{\@setfontsize\miniscule{5}{6}}% \tiny: 6/7
\or% 12pt
  \newcommand{\miniscule}{\@setfontsize\miniscule{5}{6}}% \tiny: 6/7
\fi
\makeatother

\DeclareRobustCommand{\optbar}[1]{\shortstack{{\miniscule (\rule[.5ex]{1.25em}{.18mm})}
  \\ [-.7ex] $#1$}}

\def\g      {{\ensuremath{\Pgamma}}\xspace}

\def\W      {{\ensuremath{\PW}}\xspace}

\def\squark    {{\ensuremath{\Ps}}\xspace}

\def\cquark    {{\ensuremath{\Pc}}\xspace}

\def\bquark    {{\ensuremath{\Pb}}\xspace}

\def\pion   {{\ensuremath{\Ppi}}\xspace}
\def\piz    {{\ensuremath{\pion^0}}\xspace}

\def\pip    {{\ensuremath{\pion^+}}\xspace}
\def\pim    {{\ensuremath{\pion^-}}\xspace}
\def\pipm   {{\ensuremath{\pion^\pm}}\xspace}
\def\pimp   {{\ensuremath{\pion^\mp}}\xspace}

\def\rhomeson {{\ensuremath{\Prho}}\xspace}

\def\rhop     {{\ensuremath{\rhomeson^+}}\xspace}

\def\rhopm    {{\ensuremath{\rhomeson^\pm}}\xspace}

\def\kaon    {{\ensuremath{\PK}}\xspace}
  \def\Kbar    {{\kern 0.2em\overline{\kern -0.2em \PK}{}}\xspace}

\def\KorKbar    {\kern 0.18em\optbar{\kern -0.18em K}{}\xspace}

\def\Kp      {{\ensuremath{\kaon^+}}\xspace}
\def\Km      {{\ensuremath{\kaon^-}}\xspace}
\def\Kpm     {{\ensuremath{\kaon^\pm}}\xspace}
\def\Kmp     {{\ensuremath{\kaon^\mp}}\xspace}

  \def\Dbar    {{\kern 0.2em\overline{\kern -0.2em \PD}{}}\xspace}
\def\D       {{\ensuremath{\PD}}\xspace}

\def\DorDbar    {\kern 0.18em\optbar{\kern -0.18em D}{}\xspace}
\def\Dz      {{\ensuremath{\D^0}}\xspace}
\def\Dzb     {{\ensuremath{\Dbar{}^0}}\xspace}

\def\Dstarp  {{\ensuremath{\D^{*+}}}\xspace}

\def\Dsm     {{\ensuremath{\D^-_\squark}}\xspace}

\def\Dsmp    {{\ensuremath{\D^{\mp}_\squark}}\xspace}

\def\Dssm    {{\ensuremath{\D^{*-}_\squark}}\xspace}

\def\Dssmp   {{\ensuremath{\D^{*\mp}_\squark}}\xspace}

\def\B       {{\ensuremath{\PB}}\xspace}
\def\Bbar    {{\ensuremath{\kern 0.18em\overline{\kern -0.18em \PB}{}}}\xspace}

\def\BorBbar    {\kern 0.18em\optbar{\kern -0.18em B}{}\xspace}
\def\Bz      {{\ensuremath{\B^0}}\xspace}

\def\Bu      {{\ensuremath{\B^+}}\xspace}

\def\Bp      {{\ensuremath{\Bu}}\xspace}

\def\Bd      {{\ensuremath{\B^0}}\xspace}
\def\Bs      {{\ensuremath{\B^0_\squark}}\xspace}
\def\Bsb     {{\ensuremath{\Bbar{}^0_\squark}}\xspace}

  \def\Y#1S{\ensuremath{\PUpsilon{(#1S)}}\xspace}% no space before {...}!

\def\proton      {{\ensuremath{\Pp}}\xspace}

\def\Lbar        {{\ensuremath{\kern 0.1em\overline{\kern -0.1em\PLambda}}}\xspace}
\def\LorLbar    {\kern 0.18em\optbar{\kern -0.18em \PLambda}{}\xspace}

\def\BF         {{\ensuremath{\cal B}}\xspace}

\def\BR         {\BF}
\newcommand{\decay}[2]{\ensuremath{#1\!\to #2}\xspace}         % {\Pa}{\Pb \Pc}

\def\to                 {\ensuremath{\rightarrow}\xspace}

\def\CP                {{\ensuremath{C\!P}}\xspace}

\def\AT#1     {\ensuremath{A_{\mathrm{T}}^{#1}}\xspace}           % 2

\def\C#1      {\ensuremath{\mathcal{C}_{#1}}\xspace}                       % 9
\def\Cp#1     {\ensuremath{\mathcal{C}_{#1}^{'}}\xspace}                    % 7
\def\Ceff#1   {\ensuremath{\mathcal{C}_{#1}^{\mathrm{(eff)}}}\xspace}        % 9  
\def\Cpeff#1  {\ensuremath{\mathcal{C}_{#1}^{'\mathrm{(eff)}}}\xspace}       % 7
\def\Ope#1    {\ensuremath{\mathcal{O}_{#1}}\xspace}                       % 2
\def\Opep#1   {\ensuremath{\mathcal{O}_{#1}^{'}}\xspace}                    % 7

\newcommand{\tev}{\ifthenelse{\boolean{inbibliography}}{\ensuremath{~T\kern -0.05em eV}\xspace}{\ensuremath{\mathrm{\,Te\kern -0.1em V}}}\xspace}
\newcommand{\gev}{\ensuremath{\mathrm{\,Ge\kern -0.1em V}}\xspace}
\newcommand{\mev}{\ensuremath{\mathrm{\,Me\kern -0.1em V}}\xspace}
\newcommand{\kev}{\ensuremath{\mathrm{\,ke\kern -0.1em V}}\xspace}
\newcommand{\ev}{\ensuremath{\mathrm{\,e\kern -0.1em V}}\xspace}
\newcommand{\gevc}{\ensuremath{{\mathrm{\,Ge\kern -0.1em V\!/}c}}\xspace}
\newcommand{\mevc}{\ensuremath{{\mathrm{\,Me\kern -0.1em V\!/}c}}\xspace}
\newcommand{\gevcc}{\ensuremath{{\mathrm{\,Ge\kern -0.1em V\!/}c^2}}\xspace}
\newcommand{\gevgevcccc}{\ensuremath{{\mathrm{\,Ge\kern -0.1em V^2\!/}c^4}}\xspace}
\newcommand{\mevcc}{\ensuremath{{\mathrm{\,Me\kern -0.1em V\!/}c^2}}\xspace}

\def\mum  {\ensuremath{{\,\upmu\rm m}}\xspace}

\def\invfb   {\ensuremath{\mbox{\,fb}^{-1}}\xspace}

\newcommand{\chisq}{\ensuremath{\chi^2}\xspace}

\newcommand{\chisqip}{\ensuremath{\chi^2_{\rm IP}}\xspace}

\def\gsim{{~\raise.15em\hbox{$>$}\kern-.85em
          \lower.35em\hbox{$\sim$}~}\xspace}
\def\lsim{{~\raise.15em\hbox{$<$}\kern-.85em
          \lower.35em\hbox{$\sim$}~}\xspace}

\def\sPlot{\mbox{\em sPlot}\xspace}
\def\sqs   {\ensuremath{\protect\sqrt{s}}\xspace}

\def\ptot       {\mbox{$p$}\xspace}
\def\pt         {\mbox{$p_{\rm T}$}\xspace}

\def\evtgen     {\mbox{\textsc{EvtGen}}\xspace}

\def\geant      {\mbox{\textsc{Geant4}}\xspace}

\def\photos     {\mbox{\textsc{Photos}}\xspace}

\def\pythia     {\mbox{\textsc{Pythia}}\xspace}

\def\tell1  {TELL1\xspace}
\def\ukl1   {UKL1\xspace}

\newcommand{\mytev}{\ensuremath{\mathrm{\,Te\kern -0.1em V}}\xspace} %to make TeV work in table captions

\newcommand{\p}{\ensuremath{p}\xspace}

\newcommand{\BstoDsK}{\decay{\Bs}{\Dsmp\Kpm}}

\newcommand{\BstoDspi}{\decay{\Bs}{\Dsm\pip}}

\newcommand{\BsDsPi}{\BstoDspi}
\newcommand{\BsDsK}{\BstoDsK}

\usepackage{longtable}
\usepackage{cite}
\usepackage{mciteplus}
\mciteErrorOnUnknownfalse

\begin{document}

% \linenumbers

\newcommand{\DeltaM}{$\Delta_{M}$\xspace}

\renewcommand{\thefootnote}{\fnsymbol{footnote}}
\setcounter{footnote}{1}

\begin{titlepage}
\pagenumbering{roman}

\vspace*{-1.5cm}
\centerline{\large EUROPEAN ORGANIZATION FOR NUCLEAR RESEARCH (CERN)}
\vspace*{1.5cm}
\hspace*{-0.5cm}
\begin{tabular*}{\linewidth}{lc@{\extracolsep{\fill}}r}
\ifthenelse{\boolean{pdflatex}}
{\vspace*{-2.7cm}\mbox{\!\!\!\includegraphics[width=.14\textwidth]{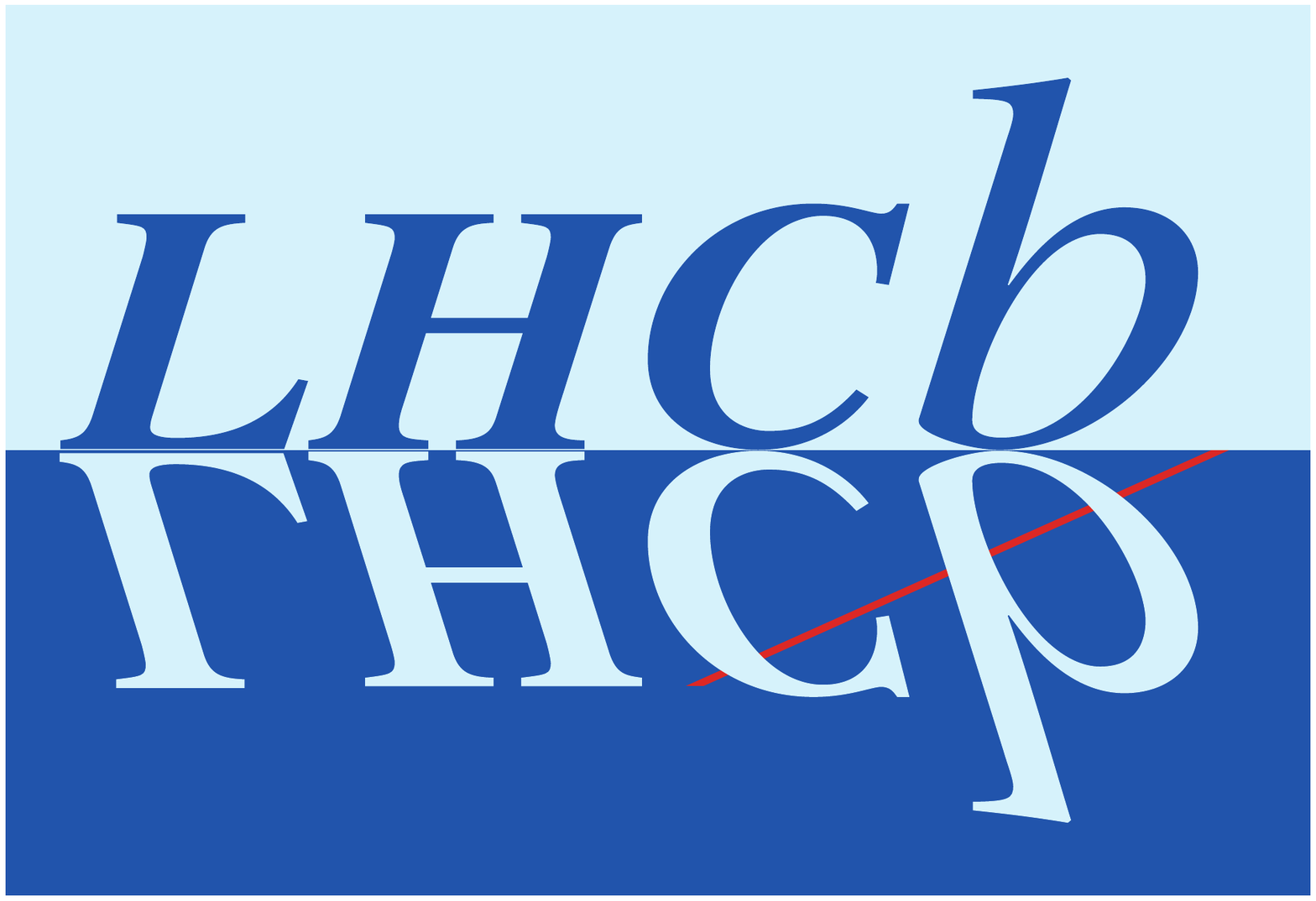}} & &}%
{\vspace*{-1.2cm}\mbox{\!\!\!\includegraphics[width=.12\textwidth]{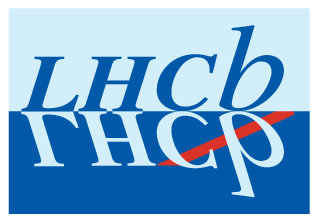}} & &}%
\\
 & & CERN-PH-EP-2015-081 \\ 
 & & LHCb-PAPER-2015-008 \\
 & & \today \\
 & & \\
\end{tabular*}

\vspace*{1.0cm}

{\bf\boldmath\huge
\begin{center}
First observation and measurement \\
of the branching fraction for the \\
decay \Bs $\to$ \Dssmp \Kpm
\end{center}
}

\vspace*{2.0cm}

\begin{center}
The LHCb collaboration\footnote{Authors are listed at the end of this paper.}
\end{center}

\vspace{\fill}

\begin{abstract}
  \noindent
  The first observation of the \Bs $\to$ \Dssmp \Kpm decay is reported
  using 3.0\invfb of proton-proton collision data collected by the \lhcb 
  experiment.
  The \Dssmp mesons are reconstructed through the decay chain
  \Dssmp $\to$ \g\Dsmp(\Kmp \Kpm \pimp).
  The branching fraction relative to that for \Bs $\to$ \Dssm \pip decays 
  is measured to be
  \begin{center}
  $\BR(\Bs $\to$ \Dssmp \Kpm)/
   \BR(\Bs $\to$ \Dssm \pip) =
  0.068 \pm 0.005 ^{+0.003}_{-0.002}$,
  \end{center}
  where the first uncertainty is statistical and the second is systematic.
  Using a recent measurement of $\BR(\Bs \to \Dssm \pip)$, 
  the absolute branching fraction of \Bs $\to$ \Dssmp \Kpm is measured as
  \begin{center}
  $\BR(\Bs \to \Dssmp \Kpm)$ =
  ( 16.3 $\pm$ 1.2 (stat) $^{+0.7}_{-0.5}$ (syst) $\pm$ 4.8 (norm) ) 
  $\times$ 10$^{-5}$,
  \end{center}
  where the third uncertainty is due to the uncertainty on the branching 
  fraction of the normalisation channel. 
\end{abstract}

\vspace*{2.0cm}

\begin{center}
Submitted to JHEP
\end{center}

\vspace{\fill}

{\footnotesize 
\centerline{\copyright~CERN on behalf of the \lhcb collaboration, license \href{http://creativecommons.org/licenses/by/4.0/}{CC-BY-4.0}.}}
\vspace*{2mm}

\end{titlepage}

\newpage
\setcounter{page}{2}
\mbox{~}

\cleardoublepage

\renewcommand{\thefootnote}{\arabic{footnote}}
\setcounter{footnote}{0}

\pagestyle{plain}
\setcounter{page}{1}
\pagenumbering{arabic}

\section{Introduction}
\label{sec:Introduction}

\begin{figure}[h]
  \begin{center}
    \includegraphics[width=.32\linewidth]{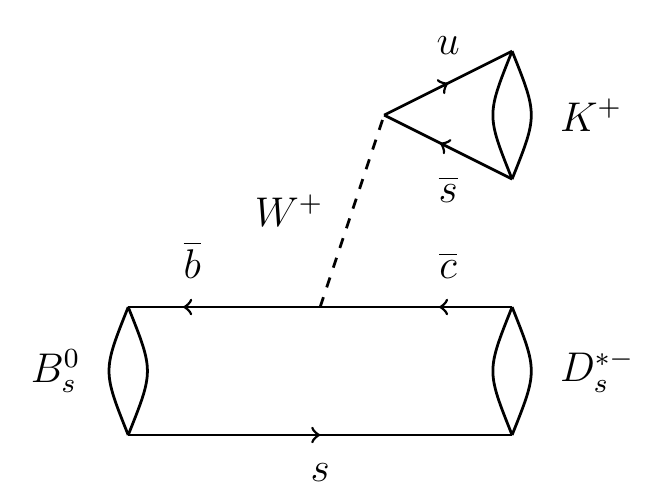}
    \hspace{2em}
    \includegraphics[width=.32\linewidth]{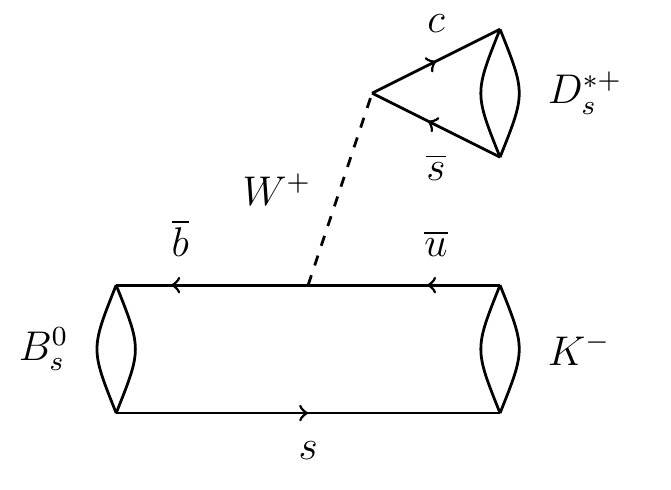}
    \includegraphics[width=.32\linewidth]{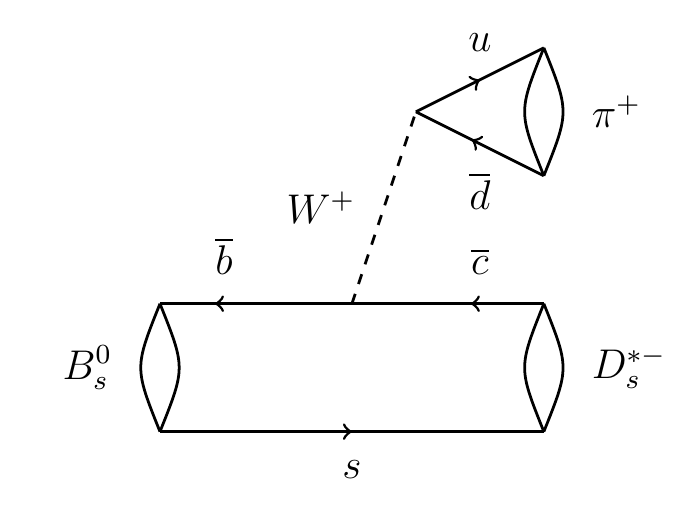}
    \hspace{3.5em}
    \includegraphics[width=.32\linewidth]{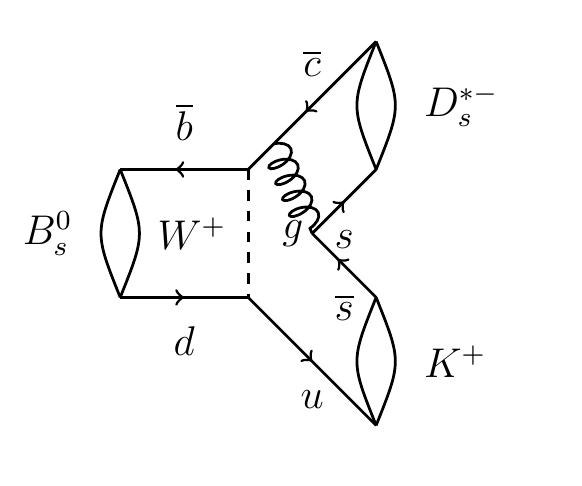}
  \end{center}
  \caption{
  \small Feynman diagrams of the processes under study. The upper diagrams
  represent the two tree topologies ($b \to c$ and $b \to u$ transitions,
  respectively) by which a \Bs meson decays into the \Dssmp \Kpm final state; 
  the lower diagrams show the tree diagram of \Bs $\to$ \Dssm \pip and the 
  \W-exchange topology of \Bs $\to$ \Dssm \Kp.}
  \label{fig:FeynmanDiagrams}
\end{figure}

The weak phase \g is one of the least well-determined CKM parameters. 
It can be measured using time-independent decay\footnote{Charge-conjugate 
states are implied throughout.} rates, such as those of \Bp $\to$ \Dzb \Kp\, 
or by time-dependent studies of \Bs $\to$ $D_s^{(*)\mp}$\Kpm 
decays~\cite{DeBruyn:2012}.
In time-dependent measurements with the decays
$B^0_{(s)} \to D_{(s)}^{(*)-} h^+$, where $h$ indicates a light meson,
the sensitivity to \g is a consequence of the interference between the
amplitudes of the $b \to u$ and $b \to c$ transitions occuring through 
$B_{(s)}^0$-$\overline{B}_{(s)}^0$ mixing.
The relevant Feynman diagrams for the \Bs system are shown in 
Fig.~\ref{fig:FeynmanDiagrams}.

The \Bs $\to$ \Dsmp \Kpm decay mode has already been used by \lhcb to determine
\g with a statistical precision of about $30^{\circ}$\cite{LHCb-PAPER-2014-038},
in an analysis based on data corresponding to an integrated luminosity of
$1\invfb$.
An attractive feature of \Bs $\to$ \Dssmp \Kpm decays is that the
theoretical formalism that relates the measured \CP asymmetries to \g 
is the same as for \Bs $\to$ \Dsmp \Kpm decays, when the angular momentum 
of the final state is taken into account in the time evolution of the 
\Bs-\Bsb decay asymmetries.

The observables of the decay \Bs $\to$ $D_s^{(*)\mp}$\Kpm can be related to 
those of \Bz $\to$ $D^{(*)-}$\pip as described in Ref.~\cite{DeBruyn:2012}
through the U--spin symmetry of strong interactions.
This opens the possibility of a combined extraction of \g.
In addition, there is a higher sensitivity to \g in
\Bs $\to$  $D_s^{(*)\mp}$\Kpm decays than in
\Bz $\to$ $D^{(*)-}$\pip decays due to the larger interference between the
$b \to u$ and $b \to c$ amplitudes in the former.

The ratio ${\cal R} \equiv \BR(\Bs \to \Dsmp \Kpm)/\BR(\Bs \to \Dsm \pip)$ has 
recently been measured by \lhcb~\cite{LHCb-PAPER-2014-064} to be 
${\cal R}= 0.0762 \pm 0.0015 \pm 0.0020$, where the first 
uncertainty is statistical and the second systematic. 
This is compatible with the predicted value of 
${\cal R}$ = $0.086^{+0.009}_{-0.007}$
from Ref.~\cite{DeBruyn:2012}, which is based on $SU(3)$ flavour symmetry
and measurements from $B$ factories. 
Under the same theoretical assumptions, the ratio 
${\cal R}^* \equiv \BR(\Bs \to \Dssmp \Kpm)/\BR(\Bs \to \Dssm \pip)$ is 
predicted to be ${\cal R}^*=0.099^{+0.030}_{-0.036}$~\cite{DeBruyn:2012} and 
it is therefore interesting to test this prediction for vector decays.

The \Bs $\to$ \Dssm \pip and \Bs $\to$ \Dssmp \Kpm decays are experimentally 
challenging for detectors operating at hadron colliders because they 
require the reconstruction of a soft photon in the \Dssm $\to$ \Dsm \g decay. 
This paper describes the reconstruction of the \Bs $\to$ \Dssm \pip
decay, previously observed by Belle~\cite{louvot:2010}, as well as
the first observation of the \Bs $\to$ \Dssmp \Kpm decay and the
measurement of ${\cal R}^*$. 
This is the first step towards a measurement of the time-dependent \CP 
asymmetry in these decays.

The $\proton\proton$ collision data used in this analysis correspond to an
integrated luminosity of $3.0\invfb$, of which $1.0\invfb$ were collected by
\lhcb in 2011 at a centre-of-mass energy of $\sqs=7\tev$, and the remaining
$2.0\invfb$ in 2012 at $\sqs=8\tev$.

The ratio of branching fractions for the decays \Bs $\to$ \Dssmp \Kpm to 
\Bs $\to$ \Dssm \pip is evaluated according to
\begin{equation}
{\cal R}^* =
    \dfrac{N_{\Kpm}}{N_{\pip}}
    \dfrac{\varepsilon_{\pip}}{\varepsilon_{\Kpm}},
\label{eq:BRratio}
\end{equation}
where $\varepsilon_{X}$ and $N_{X}$ are the overall reconstruction efficiency 
and the observed yield, respectively, of the decay mode, and $X$ represents 
either a kaon or a pion (the ``bachelor'' hadron) that accompanies the \Dssm 
in the final state.

\section{\lhcb detector}

The \lhcb detector~\cite{Alves:2008zz,LHCb-DP-2014-002} is a single-arm 
forward spectrometer covering the \mbox{pseudorapidity} range $2<\eta <5$,
designed for the study of particles containing \bquark or \cquark
quarks. The detector includes a high-precision tracking system consisting of 
a silicon-strip vertex detector surrounding the $pp$ interaction region, a 
large-area silicon-strip detector located upstream of a dipole magnet with a 
bending power of about $4{\rm\,Tm}$, and three stations of silicon-strip 
detectors and straw drift tubes placed downstream of the magnet.
The tracking system provides a measurement of momentum, \ptot, of charged 
particles with a relative uncertainty that varies from 0.5\% at low momentum 
to 1.0\% at 200\gevc.
The minimum distance of a track to a primary vertex, the impact parameter, is 
measured with a resolution of $(15+29/\pt)\mum$, where \pt is the component 
of the momentum transverse to the beam, in \gevc.
Different types of charged hadrons are distinguished using information
from two ring-imaging Cherenkov detectors. 
Photons, electrons and hadrons are identified by a calorimeter system 
consisting of scintillating-pad and preshower detectors, an electromagnetic
calorimeter and a hadronic calorimeter. Muons are identified by a
system composed of alternating layers of iron and multiwire
proportional chambers.

The online event selection is performed by a trigger which consists of a 
hardware stage, based on information from the calorimeter and muon systems, 
followed by a software stage, which applies a full event reconstruction.
At the hardware trigger stage, events are required to have a muon with high
\pt or a hadron, photon or electron with high transverse energy in the
calorimeters.
For hadrons, the transverse energy threshold is 3.5 GeV.
The software trigger requires a two-, three- or
four-track secondary vertex with a significant displacement from the primary
$pp$ interaction vertices~(PVs). At least one charged particle must have a
transverse momentum $\pt > 1.7 \gevc$ and be inconsistent with originating from
a PV. A multivariate algorithm~\cite{BBDT} is used for the identification of
secondary vertices consistent with the decay of a \bquark hadron.
The \pt of the photon from \Dssm decay is too low to contribute to the trigger
decision.

In the simulation, $pp$ collisions are generated using
\pythia~\cite{Sjostrand:2006za,*Sjostrand:2007gs} 
with a specific \lhcb configuration~\cite{LHCb-PROC-2010-056}.  Decays of 
hadronic particles are described by \evtgen~\cite{Lange:2001uf}, in which 
final-state radiation is generated using \photos~\cite{Golonka:2005pn}. The
interaction of the generated particles with the detector, and its response,
are implemented using the \geant 
toolkit~\cite{Allison:2006ve, *Agostinelli:2002hh} 
as described in Ref.~\cite{LHCb-PROC-2011-006}.

\section{Event selection}

Candidate \Bs~mesons are reconstructed by combining a \Dssm~candidate
with an additional pion or kaon of opposite charge.
The preselection and selection for the two decays analysed for the 
measurement of ${\cal R}^*$ differ only by the particle identification 
(PID)~\cite{LHCb-DP-2012-003} requirements imposed on the bachelor 
tracks.
The \Dssm and \Dsm candidates are reconstructed in the \Dsm \g and
\Km \Kp \pim decay modes, respectively.
Each of the three \Dsm daughters tracks is required to have a good track
quality, momentum \p $>$ 1000\mevc, transverse momentum \pt $>$ 100\mevc 
and a large impact parameter with respect to any PV.
More stringent requirements are imposed for bachelor tracks, namely 
\p $>$ 5000\mevc and \pt $>$ 500\mevc.
A good quality secondary vertex is required for the resulting \Dsm-bachelor 
combination.
Photons are identified using energy deposits in the electromagnetic 
calorimeter that are not associated with any track in the tracking system.
Due to the small difference between the masses of the \Dssm and \Dsm mesons,
called \DeltaM in the following, the photons from the \Dssm decay have an 
average transverse energy of a few hundred \mevcc.
A cut on a photon confidence level variable is used to suppress background 
events from hadrons, electrons and \piz decays~\cite{LHCb-DP-2014-002}. 
This confidence level variable takes into account the
expected absence of matching between the calorimeter cluster and any
track, the energy recorded in the preshower detector and the topology of the 
energy deposit in the electromagnetic and hadronic calorimeters.

Additional preselection requirements are applied to cope with a large 
background mainly due to genuine photons that are not \Dssm decay products,
or hadrons that are misidentified as photons.
The reconstructed mass of the \Dsm candidate and the reconstructed \DeltaM 
value are required to be in a $\pm$ 20\mevcc window around their known
values~\cite{PDG2014}.
The $B^0_{(s)}$~\to \Dsm \Kp (\pip) decays are vetoed by a cut on the invariant
mass of the \Dsm \Kp(\pip) system.
PID requirements are applied to all final-state hadrons.
Finally, the maximum distance in the $\eta$--$\varphi$ plane between the 
\Dsm and the photon is required to satisfy 
$\sqrt{\Delta\eta^2+\Delta\varphi^2} <1$,
where $\Delta\eta$ ($\Delta\varphi$) is the pseudo-rapidity (azimuthal angle) 
distance between the corresponding candidates.

To further reduce the combinatorial background while preserving a high
signal efficiency, a multivariate approach is used. 
This follows closely the selection based on a boosted decision tree 
(BDT)~\cite{Breiman,AdaBoost} used in the measurement of the ratio of 
\BsDsK to \BsDsPi branching fractions~\cite{LHCb-PAPER-2014-064}.
The algorithm is trained with simulated \Bs $\to$ \Dssm \pip events as signal,
and candidates in data with an invariant mass greater than 5500~\mevcc as
background.
The five variables with the highest discriminating power are found to be the 
\Bs transverse flight distance, the photon transverse momentum,
the \chisqip of the \Bs candidate (where \chisqip
is defined as the difference in \chisq of the associated PV, reconstructed
with and without the considered particle),
the angle between the \Bs momentum vector and the vector connecting its 
production and decay vertices, and the transverse momentum of
the bachelor particle. Eight additional variables, among them the transverse
momenta of the remaining final-state particles, are also used.
The trained algorithm is then applied to both the \Bs $\to$ \Dssmp \Kpm and 
\Bs $\to$ \Dssm \pip decays.

\begin{figure}
\begin{center}
\includegraphics[width=16cm]{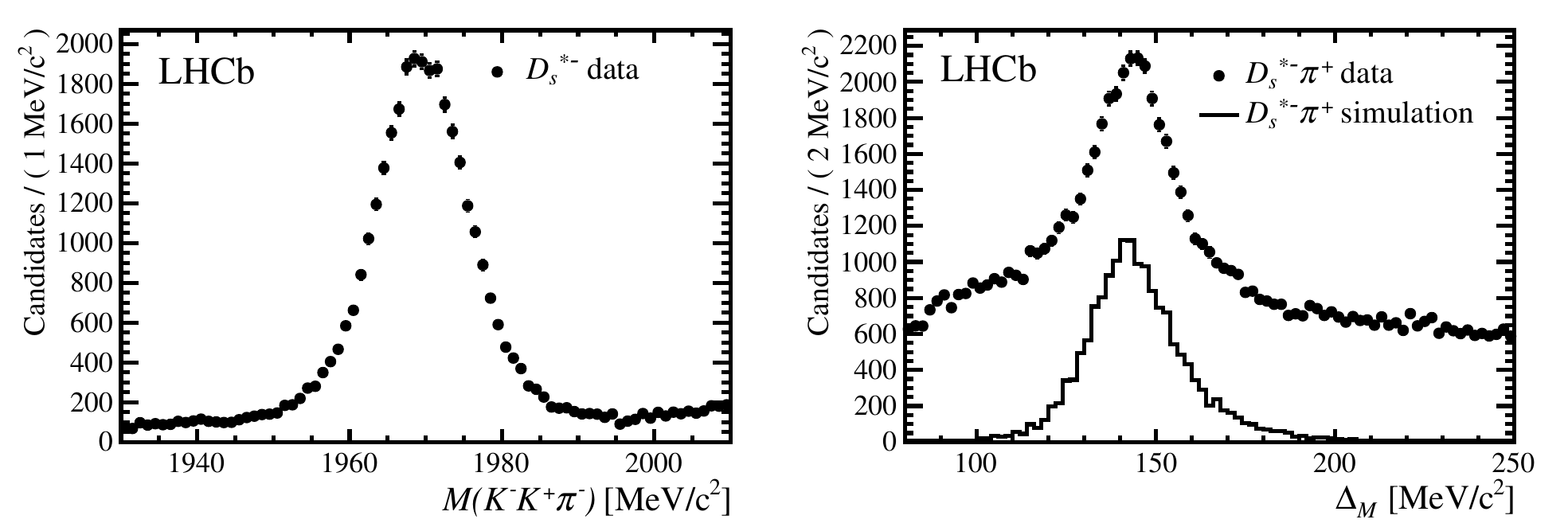}
\caption{(left) The \Km \Kp \pim invariant mass and (right) mass difference 
\DeltaM of the \Bs $\to$ \Dssm \pip candidates.
The points represent data. On the right plot the solid line represents
the signal expected from the simulations.}
\label{fig:DsDeltaM}
\end{center}
\end{figure}

The $M(\Km\Kp\pim)$ and \DeltaM invariant mass distributions, as obtained from 
the decay mode \Bs $\to$ \Dssm \pip, are shown in Fig.~\ref{fig:DsDeltaM}. 
These distributions have been obtained with all of the analysis requirements 
applied except that on the plotted variable.
In both cases the \Bs invariant mass is restricted to a $\pm$ 70\mevcc 
region around the known mass.
A prominent peaking structure is observed in the \DeltaM distribution around 
145\mevcc, due to the radiative \Dssm to \Dsm decay.

\section{Signal yields}

The signal yields are obtained using unbinned maximum likelihood fits to 
the \Bs candidate invariant mass distributions and are performed separately
for \Bs $\to$ \Dssm \pip and \Bs~$\to$~\Dssmp \Kpm decays.

The signal shapes are parametrised by a double-sided Crystal Ball (CB) 
function~\cite{Skwarnicki:1986xj}, which consists of a central Gaussian part,
with mean and width as parameters, and power-law tails on both lower and 
upper sides, to account for energy loss due to final-state radiation and 
detector resolution effects.
The two mean values are constrained to be equal.
When fitting the \Dssm \pip and \Dssmp \Kpm simulated mass distributions 
all parameters are floated.
When fitting data, the power-law tails parameters are fixed to the result
of the fit to the corresponding simulation.
Furthermore, both widths of the CB are set to those obtained from the signal 
simulation, scaled by a variable parameter in the fit to allow for differences 
in the mass resolution between data and simulation.
The common mean of the double-sided CB is allowed to vary.

Three background categories are identified. Partially reconstructed background
decays are due to \Bs decay modes that are similar to signal but with at least 
one additional photon, as for example in the case of the 
\Bs $\to$ \Dssmp \rhopm 
decays with \rhopm $\to$ \piz ($\to$ \g \g) \pipm.
Fully reconstructed background events are due to \Bd decays to the same final 
states as the \Bs signal, \Dssm \pip and \Dssmp \Kpm.
The \Bs $\to$ \Dssm \pip decays gives rise to a peak in the 
\Bs~$\to$~\Dssmp \Kpm decay mode when the \pip is misidentified as a \Kp, a 
cross feed contribution. 
The cross feed due to \Kpm to \pipm misidentification is negligible.
Finally, a combinatorial background, where a genuine \Dsm meson is combined 
with a random (or fake) photon and a random bachelor track, can also
contribute. 

The number of partially and fully reconstructed background components is
different for each of the two final states.
The invariant mass shapes for these backgrounds are obtained from 
simulation and are represented in the fit as non-parametric probability
density functions (PDFs).
The yields of these background components are free parameters in the fit, 
with the exception of the \Dssm \pip, \Dsm \rhop and \Dssm \rhop
contributions in the \Dssmp \Kpm fit. 
The size of the \Dssm \pip cross feed is calculated from the 
\Dssm \pip yield and the \pion to \kaon misidentification probability. 
The \Dsm \rhop and \Dssm \rhop contributions are determined in a similar 
manner, summed and fixed in the fit.

To model the combinatorial background a non-parametric PDF is used.
This is obtained from the events of the \DeltaM sideband in the 
interval [185,205]\mevcc, with all other cuts unchanged.

\begin{figure}[htbp]
\begin{center}
\includegraphics[width=12.cm]{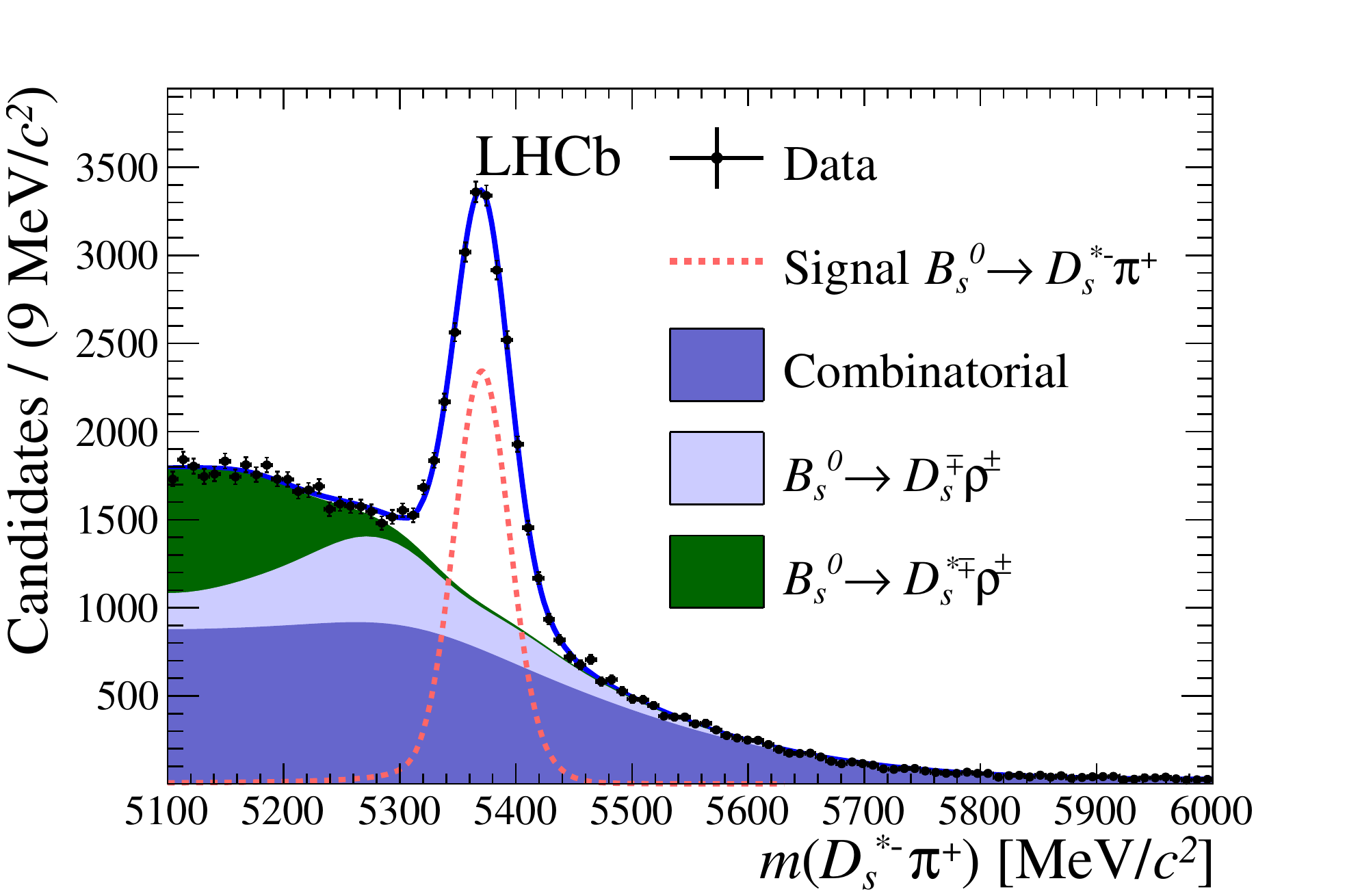}
\includegraphics[width=12.cm]{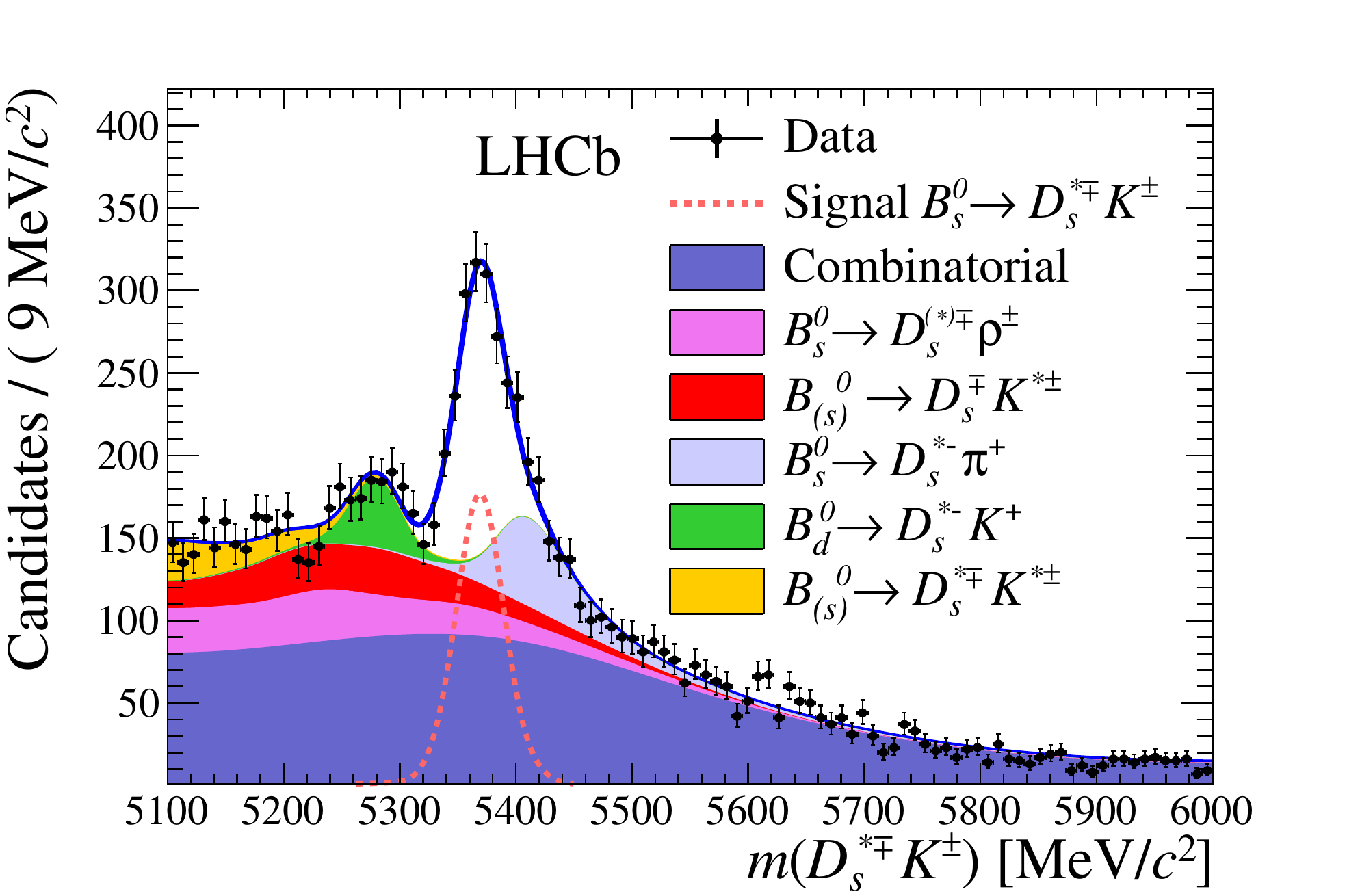}
\end{center}
\caption{Invariant mass distribution of (top) \Bs $\to$ \Dssm \pip and 
(bottom) \Bs $\to$ \Dssmp \Kpm candidates with fit results superimposed.
The fitted signal corresponding to the first observation of 
\Bs $\to$ \Dssmp \Kpm is shown by the dotted line in the lower plot.}
\label{fig:fitPiK}
\end{figure}

The results of the fitting procedure applied to the two considered decay
modes are shown in Fig.~\ref{fig:fitPiK}.
The fitted yields are 16$\,$513 $\pm$ 227 and 1025 $\pm$ 71 for the
\Bs $\to$ \Dssm \pip and \Bs $\to$ \Dssmp \Kpm cases, respectively.
When the $\chi^2$ test is applied to gauge the quality of the fits, the
latter fit has a $\chi^2$ value of 88.5 for 100 bins and 7 free parameters, 
the quality of the former fit is equally good. 

\begin{figure}
\begin{center}
\includegraphics[width=16cm]{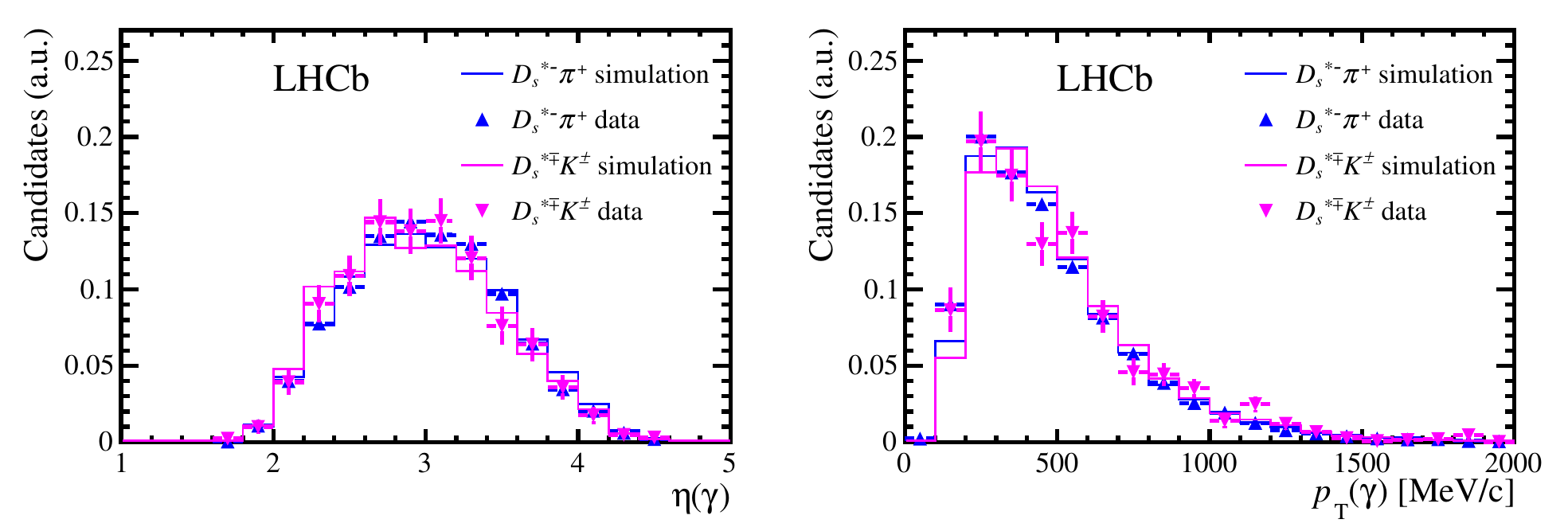}
\caption{Distributions of (left) $\eta$ and (right) \pt of the 
photons for the \Dssm \pip (blue) and \Dssmp \Kmp (magenta) decays.
Data, background-subtracted using the \sPlot method, are represented 
by points, and simulations by solid lines.}
\label{fig:photon}
\end{center}
\end{figure}

One of the distinctive features of the present analysis is the reconstruction 
of the decay mode \Dssm $\to$ \Dsm \g at a hadron collider.
The background-subtracted $\eta$ and \pt distributions of these photons
have been obtained using the invariant mass fit results described above and 
the \sPlot~\cite{Pivk:2004ty} method.
These measured distributions are compared to the predictions of the simulation 
in Fig.~\ref{fig:photon}.
It is noted that most of the measured photons are very soft, with the average 
\pt well below 1\gevc.

\section{Systematic uncertainties}

Potential systematic uncertainties on ${\cal R}^*$ are those due to the 
background modelling and the analysis selections, including the BDT 
and the PID cuts.
Their effects are shown in Table~\ref{tab:sys} as relative variations 
of the final result, with their sum in quadrature assigned as the overall 
systematic uncertainty. 
The order in which the systematic uncertainties are described in the 
following text corresponds to successive rows in Table~\ref{tab:sys}.

Combinatorial background modelling uncertainties are studied by varying
the default \DeltaM range used for the combinatorial background 
determination, [185,205]\mevcc, to [205,225] and [225,245]\mevcc.
An alternative modelling of this background, using a parametric shape 
obtained from the \Dsm mass sidebands, is also tested.
Finally, the statistical uncertainty due to the number of events 
in the range [185,205]\mevcc is evaluated using the bootstrap 
technique~\cite{Efron:1979,Narsky2013}.
The corresponding uncertainty is taken to be the largest spread among
the four differents checks.

The uncertainty due to the finite size of the simulated samples used
to study the partially reconstructed backgrounds is studied using the 
bootstrap technique.

The uncertainties due to the \Dssm \pip cross feed and the \Dsm \rhop and 
\Dssm \rhop contributions to the \Dssmp \Kpm fit are estimated by varying 
their expected yields.
For the \Dssm \pip cross feed the $\pm 1 \sigma$ variation is obtained 
using the \Dssm \pip fit results.
In the \Dsm \rhop and \Dssm \rhop cases the branching ratio uncertainties
and photon kinematic distributions are different from the \Dssm \pip ones
so the uncertainty in the yields are large.
These yields are conservatively varied by $\pm 50$ \%. 
The observed differences in the final result are assigned as the systematic 
uncertainties associated with these sources.

The systematic uncertainty associated with the BDT is studied by reweighting 
the simulation to improve the agreement with data~\cite{LHCb-PAPER-2014-064}.
 
The \pion and \kaon PID efficiencies 
used for the bachelor track have been extracted from a \Dstarp~$\to$~\Dz \pip 
calibration sample and parametrized as a function of several kinematic 
quantities of these tracks.
The uncertainties in this procedure, propagated to the final result, lead to 
the PID systematic uncertainty.

The systematic uncertainty from the hardware trigger efficiency arises from 
differences in the pion and kaon trigger efficiencies which are not reproduced 
in the simulation~\cite{LHCb-PUB-2011-026}. The uncertainty is scaled with the 
fraction of events where a signal track was responsible for triggering.

\begin{table}[h]
\begin{center}
\caption{Estimated systematic uncertainties on ${\cal R}^*$.}
\label{tab:sys}
\renewcommand{\arraystretch}{1.2}
\begin{tabular}{cc}
\hline
source & relative variation (\%) \\
\hline
combinatorial background & $^{+ 4.7}_{- 2.2}$ \\
simulation sample size & {\small $\pm$1.4} \\
$D_s^{*-} \pi^+$ cross feed & {\small $\pm$0.8} \\
$D_s^{(*)-} \rho^+$ ``cross feed'' & $^{+0}_{-1.6}$ \\
BDT  & {\small $\pm$0.5} \\
PID uncertainties & {\small $\pm$1.0} \\
hardware trigger & {\small $\pm$1.0} \\
\hline
total & $^{+ 5.2}_{- 3.5}$ \\
\hline
\end{tabular}
\end{center}
\end{table}

\section{Results}

The ratio of branching fractions, measured in this analysis for the first
time, is
\begin{center}
${\cal R}^*$
$\equiv$ $\BR(\Bs \to \Dssmp \Kpm)/\BR(\Bs \to \Dssm \pip)$
= 0.068 $\pm$ 0.005 (stat) $^{+0.003}_{-0.002}$ (syst),
\end{center}
where the overall systematic uncertainty is mainly due to the uncertainty
on the combinatorial background estimate.
The result for ${\cal R}^*$ differs from the uncorrected \Bs $\to$ \Dssmp \Kpm 
to \Bs $\to$ \Dssm \pip events ratio by a factor depending on the 
simulation and the PID efficiencies.
This factor is determined to be $1.095 \pm 0.016$ and is dominated by 
the \kaon to \pion PID efficiency ratio.

The measured value of ${\cal R}^*$ is consistent with the theoretical
prediction of ${\cal R}^*=0.099^{+0.030}_{-0.036}$~\cite{DeBruyn:2012},
within the very large uncertainty of the latter.
The theory is found to provide a good description of the measurements for 
both ${\cal R}^*$ and ${\cal R}$\cite{LHCb-PAPER-2014-064}.
Other theoretical predictions of ${\cal R}^*$ have been published 
in Refs.~\cite{blasi,li,azizi,chen,faustov}.

Combining the measured value of ${\cal R}^*$ with the value of 
$\BR(\Bs \to \Dssm \pip)$ obtained by Belle~\cite{louvot:2010} leads to
\begin{center}
$\BR(\Bs \to \Dssmp \Kpm)$ = ( 16.3 $\pm$ 1.2 (stat) 
$^{+0.7}_{-0.5}$ (syst) $\pm$ 4.8 (norm) ) $\times$ 10$^{-5}$, \\ 
\end{center}
where the uncertainties are statistical, systematic and due to the uncertainty 
on $\BR(\Bs \to \Dssm \pip)$.

\section*{Acknowledgements}

\noindent We express our gratitude to our colleagues in the CERN
accelerator departments for the excellent performance of the LHC. We
thank the technical and administrative staff at the LHCb
institutes. We acknowledge support from CERN and from the national
agencies: CAPES, CNPq, FAPERJ and FINEP (Brazil); NSFC (China);
CNRS/IN2P3 (France); BMBF, DFG, HGF and MPG (Germany); INFN (Italy); 
FOM and NWO (The Netherlands); MNiSW and NCN (Poland); MEN/IFA (Romania); 
MinES and FANO (Russia); MinECo (Spain); SNSF and SER (Switzerland); 
NASU (Ukraine); STFC (United Kingdom); NSF (USA).
The Tier1 computing centres are supported by IN2P3 (France), KIT and BMBF 
(Germany), INFN (Italy), NWO and SURF (The Netherlands), PIC (Spain), GridPP 
(United Kingdom).
We are indebted to the communities behind the multiple open 
source software packages on which we depend. We are also thankful for the 
computing resources and the access to software R\&D tools provided by Yandex LLC (Russia).
Individual groups or members have received support from 
EPLANET, Marie Sk\l{}odowska-Curie Actions and ERC (European Union), 
Conseil g\'{e}n\'{e}ral de Haute-Savoie, Labex ENIGMASS and OCEVU, 
R\'{e}gion Auvergne (France), RFBR (Russia), XuntaGal and GENCAT (Spain), Royal Society and Royal
Commission for the Exhibition of 1851 (United Kingdom).

\setboolean{inbibliography}{true}
\bibliographystyle{LHCb}
\bibliography{main,LHCb-PAPER,LHCb-CONF,LHCb-DP,LHCb-TDR}

\newpage
%%%%%%%%%%%%%%%%%%%%%%%%%%%%%%%%%%%%%%%%%%
\centerline{\large\bf LHCb collaboration}
\begin{flushleft}
\small
R.~Aaij$^{41}$, 
B.~Adeva$^{37}$, 
M.~Adinolfi$^{46}$, 
A.~Affolder$^{52}$, 
Z.~Ajaltouni$^{5}$, 
S.~Akar$^{6}$, 
J.~Albrecht$^{9}$, 
F.~Alessio$^{38}$, 
M.~Alexander$^{51}$, 
S.~Ali$^{41}$
% ...

{\footnotesize \it
% ...
$ ^{5}$Clermont Universit\'{e}, Universit\'{e} Blaise Pascal, CNRS/IN2P3, LPC, Clermont-Ferrand, France\\
$ ^{6}$CPPM, Aix-Marseille Universit\'{e}, CNRS/IN2P3, Marseille, France\\
% ...
$ ^{9}$Fakult\"{a}t Physik, Technische Universit\"{a}t Dortmund, Dortmund, Germany\\
% ...
$ ^{37}$Universidad de Santiago de Compostela, Santiago de Compostela, Spain\\
$ ^{38}$European Organization for Nuclear Research (CERN), Geneva, Switzerland\\
% ...
$ ^{41}$Nikhef National Institute for Subatomic Physics, Amsterdam, The Netherlands\\
% ...
$ ^{46}$H.H. Wills Physics Laboratory, University of Bristol, Bristol, United Kingdom\\
% ...
$ ^{51}$School of Physics and Astronomy, University of Glasgow, Glasgow, United Kingdom\\
$ ^{52}$Oliver Lodge Laboratory, University of Liverpool, Liverpool, United Kingdom\\
% ...
}
\end{flushleft}
%%%%%%%%%%%%%%%%%%%%%%%%%%%%%%%%%%%%%%%%%%

% \newpage
% \include{tikz}

\end{document}